\documentclass[manuscript]{aastex}

\slugcomment{}

\shorttitle{IR spectra of M~giants}
\shortauthors{Rich, Origlia \& Valenti}

\begin{document}


\title{Detailed abundances for M giants in two inner bulge fields \\
   from Infrared Spectroscopy\footnote{Data presented herein were obtained
   at the W.~M.~Keck Observatory, which is operated as a scientific partnership
   among the California Institute of Technology, the University of California, and
   the National Aeronautics and Space Administration. The Observatory was made 
   possible by the generous financial support of the W.~M.~Keck Foundation.}}


\author{R.~M. Rich}
\affil{Department of Physics and Astronomy, University of California at Los Angeles, Los Angeles, CA 90095 - 1562, US}
\email{rmr@astro.ucla.edu}

\author{L. Origlia}
\affil{INAF - Osservatorio Astronomico di Bologna, via Ranzani 1, I-40127 Bologna, Italy}
\email{livia.origlia@oabo.inaf.it}

\and

\author{E. Valenti}
\affil{European Southern Observatory, Karl Schwarzschild\--Stra\ss e 2, 
D\--85748 Garching bei M\"{u}nchen, Germany.}
\email{evalenti@eso.org}


\begin{abstract}

We report abundance analysis for 30 M giant stars in two inner Galactic
bulge fields at $(l,b)=(0^\circ,-1.75^\circ)$ and at $(l,b)=(1^\circ,-2.65^\circ)$, based on $R=25,000$ infrared spectroscopy  ${\rm (1.5-1.8 ~\mu m)}$ using NIRSPEC at the Keck II telescope.  We find iron abundances of
$\rm \langle [Fe/H] \rangle = -0.16 \pm 0.03$ dex
with a $1\sigma$ dispersion of $0.12\pm 0.02$ and 
$\rm \langle [Fe/H] \rangle = -0.21 \pm 0.02$ dex,
with a $1\sigma$ dispersion of $0.09\pm 0.016$ 
for the $(l,b)=(0^\circ,-1.75^\circ)$ and $(l,b)=(1^\circ,-2.65^\circ)$ fields, respectively.  In agreement with all prior studies, we find enhanced $\rm [\alpha/Fe]$ of +0.3 dex.  We confirm the lack of any major vertical abundance or 
composition gradient in the innermost $\sim 600$ pc between Baade's window 
and  $\sim 150$ pc from the Galactic plane.  We also confirm that the known enhancement of alpha elements
observed between 500 and 1000 pc from the nucleus is also present over the volume of the inner bulge and may therefore be presumed to be a general characteristic
of bulge/bar stars within 1 kpc of the Galactic Center.

\end{abstract}




\section{Introduction}
\label{intro}
Even with recent observational successes, the timescales for formation and chemical enrichment of the Milky Way bulge remain a subject of debate at the several Gyr level.  The global chemistry  \citep[see e.g.][]{MWR94, fmwr07, gon11a, john11} and color-magnitude diagrams corrected for non-bulge contamination \citep{ort95,fg00,kr02,zoc03,cla08} are generally consistent with a globular cluster-aged bulge that formed early and rapidly \citep{bal07,cesc11}.  That is, bulge giants are generally alpha enhanced, which is attributed to early enrichment from core collapse SNe \citep{wst89,mcw97}.  On the other hand, microlensing studies \citep{ben11} find $\sim 25\%$ of microlensed metal rich dwarfs are as much as 5 Gyr younger than the canonical globular cluster age.   Using gravity and effective temperature measured from the spectra, the stars are placed on isochrones appropriate for the derived [Fe/H] and $[\alpha \rm /Fe]$, assuming they are at the distance of the bulge; the dissonance between the CMD and spectroscopic studies remains unsettled.

In the inner bulge, the region $\rm \sim 1^\circ = 140 pc$ from the Galactic Center, it is well established that there is a more continuous star formation history: there is a clearly detected red clump $>1$ Gyr old, but also clusters as young as a few Myr and young, massive, stars in the central pc \citep{fig04}.  However, within $\sim 300$ pc, the optical extinction increases to the point where high resolution optical spectroscopy is not currently feasible, hence the need to turn to infrared techniques if one seeks to constrain the history of chemical evolution in this region, using abundance analysis.

The last five years has seen a burgeoning number of optical and infrared spectroscopic studies at resolution $>$20,000, 
aimed at characterizing the chemical and kinematic properties of the bulge stellar populations. 
The largest recent surveys are based on high resolution optical spectroscopy of red clump stars, 
hence K giants \citep{fmwr06,zoc08,john11}.
According to these studies, there is general agreement on the bulge metallicity 
distribution spanning a fairly large range, $\rm -1.5 \le [Fe/H] \le +0.5$~dex, 
with a mean value that depends on the latitude of the observed field. 
In particular, \citet{zoc08} found an iron abundance 
gradient of $ \sim 0.6~dex~ kpc^{-1}$ along the bulge minor axis when moving inwards from a region 
at b=$-12^{\circ}$ to Baade's window at b=$-4^{\circ}$.  \citet{john11} added the abundance distribution at $b=-8^\circ$, confirming the overall trend.    Over this entire volume, the alpha elements are observed to be enhanced relative to the thin disk. The optical spectra of K giants in all of the bulge fields observed so far \citep{fmwr07,lecu07,mel08,albr10,john11,gon11a} show an overall [$\alpha$/Fe] enhancement, of $\sim$0.2\--0.4~dex, with a ``knee'' breaking toward Solar scaled alphas at  [Fe/H]$\sim -0.4$.   The exact location of this break varies slightly among different studies, although overall, the general enhancement of alpha elements relative to the thin disk supports a rapid formation scenario.

Modeling of the bulge kinematics \citep{shen10} as measured by the Bulge Radial Velocity Assay ({\it BRAVA}) \citep{richetal07,how08} project determines that for $|b|<8^\circ$, the velocity/dispersion field is fit by an N-body bar that evolved from the dynamical buckling of a disk.  Bars formed by a purely dynamical process are not expected to exhibit an abundance gradient, because acceleration should be independent of metallicity.   \citet{bab10}, \citet{hill11} and \citet{gon11a} argue that the abundance gradient arises owing to a more metal rich population that predominates at low Galactic latitude with \citet{gon11a} suggesting that it is due to a thin disk.   However, on the kpc scales of the bulge/bar, there is no evidence for a dynamically hotter classical bulge or other complexity, such as a rapidly rotating (i.e. disklike) metal rich component \citep{how08,how09,kun11}.  Further, \citet{bab10} and \citet{gon11a} argue for a metal rich population with {\it higher} velocity dispersion having {\it smaller} vertical scale height-a result that would seemingly run counter to the expected behavior of a stellar distribution function.   While one might wish to attribute the abundance gradient to the transition between the bar and a ``classical'' spheroid, the observed cylindrical rotation field characteristic of a bar (now also confirmed at b=$-6^\circ$; \citet{kun11}) extends to 1 kpc=$-8^\circ$.    One cannot attribute the peculiar hot metal rich stars at b=$-4^\circ$ to the kinematic expression of the bar: the cylindrical rotation field (modeled as due to the bar) dominates the velocity field to b=$-8^\circ$.    {\it The BRAVA kinematic survey is not compatible with a low latitude metal rich disk or bar and a high latitude ``classical'' spheroid.} Any theory for the gradient must reflect the reality of the observed global kinematics.  A vertical abundance gradient might also have arisen owing to the most metal poor (possibly oldest) stars having more time to scatter off of the bar via resonant heating \citep[see e.g.][]{pfen90}.  However, the uniformity of the $\rm [\alpha/Fe]$ trends throughout the bulge pose problems for this scenario, because the similar $\rm [\alpha/Fe]$ trends would appear to require a uniform enrichment history yet somehow leave an abundance gradient.   The gradient might also have been in place early, following the formation of a massive proto-disk, with that abundance gradient relatively unmodified by the subsequent secular evolution that produced the bar.

We emphasize that the measurement of the vertical abundance gradient cited above are based on optical spectroscopy of 
K giants in 4 outer fields along the minor axis at b=-4$^{\circ}$, -6$^{\circ}$, -8$^{\circ}$, and -12$^{\circ}$, 
with the only exception being the results of \citet{ben10,ben11} who observed microlensed dwarfs 
located at various galactic longitude and at latitudes between -2$^{\circ}$ and -4$^{\circ}$.   One other study is based
on medium resolution infrared spectroscopy, finding no gradient in fields from $-0.28^\circ$ to $-4^\circ$ \citep{ram00a}.
The bulge inner region (b$\leq 3^{\circ}$) is most seriously affected by extinction and crowding and has no prior high resolution spectroscopy except for the supergiants in the Galactic Center; it appeared to be logical to extend our
investigation into this region. 

To test whether the metallicity gradient found in the outer regions is also present in the innermost 
300~pc, we have obtained IR spectra at R=25,000 for a sample of bulge M giants in different fields 
with galactic latitude b$\leq 4^{\circ}$.  In \citet{rich05} we have reported the first detailed IR abundances for our sample of M giants in Baade's window.   

Our infrared spectroscopy has been in essentially good agreement with the optical studies, and also with subsequent IR work \citep{ryde09,ryde10} in Baade's window.  We find roughly Solar mean abundance and enhanced $\alpha$-elements, 
consistent with a predominant chemical enrichment by type II SNe on a relatively short timescale.

Very similar results were obtained when the second sample of M giants located in the innermost field at latitude 
b=-1$^{\circ}$ was analyzed \citep{rich07}. We found no evidence of any major iron abundance or 
[$\alpha$/Fe] abundance ratio gradients between the field at b=-1$^{\circ}$ and the Baade's window.

Here we present a detailed abundance analysis of an additional sample of 30 M giants observed in two additional inner bulge fields located at (l,b)=(0$^{\circ}$,-1.75$^{\circ}$) and (l,b)=(1.25$^{\circ}$,-2.65$^{\circ}$).

\section{Observations and spectral analysis}

A sample of M giants near the old red giant branch (RGB) tip \citep{fw87} in two inner bulge fields located at
(l, b)=(0$^\circ$, -1.75$^\circ$) and (l, b)=(1$^\circ$, -2.65$^\circ$) 
(hereafter F175 and F265, respectively) has been 
observed using the cross-dispersed echelle spectrograph NIRSPEC \citep{nir} at Keck~II during two observing runs
in May 2005 and May 2006. The targets have been selected by using two near\--IR 
photometric catalogs obtained from a combination of 2MASS/WIRC\footnote{The two
bulge fields have been observed with the IR camera WIRC at the Las Campanas Observatory
through the K$_s$ filter on August 2002.} data for the J and K$_s$ magnitudes, 
respectively. The derived color\--magnitude diagrams have been corrected for reddening
adopting E(B\--V)=1.2 and 0.65 (for F175 and F265, respectively). These values have been derived
by averaging the latest extinction estimates provided by the MACHO microlensing
survey \citep{redPCB03}, 2MASS data \citep{redD03}, and the OGLE survey \citep{redS04}.
In each bulge field, we obtained spectra for 15 stars with bolometric 
magnitude M$_{bol}<$-3.0 and 0.9$\leq (J\--K)_0\leq$1.4.   Considering the excellent agreement with the
bulge color-magnitude diagram and large velocity dispersion of the targets, we are confident that these are
members of the old $\sim10$ Gyr old bulge population.

Fig.~\ref{cmd} shows the 2MASS color-magnitude diagram of Baade's window, dereddened according to the prescription 
of \citet{gon11ab} and superimposed, our 30 M giants in the F175 and F265 fields as well as the other 29 giants in both
Baade's window \citep{rich05} and the b=-1$^{\circ}$ \citep{rich07} fields.  The color range of our sample is great enough to span the full range seen the CMD; we believe that our sample is unbiased with respect to metallicity.

We employed a slit width of 0.43$\arcsec$ giving an overall spectral resolution $R=25,000$, and the
standard NIRSPEC\--5 setting, which covers most of the 1.5\--1.8 $\mu$m H\--band.
The raw stellar spectra have been reduced using the REDSPEC IDL\--based package 
written at the UCLA IR Laboratory.
Each order has been sky\--subtracted by using nodding pairs and is flat\--field\--corrected.
Wavelength calibration has been performed using arc lamps and a second\--order polynomial
solution, while telluric features have been removed by using an O star featureless spectrum.
The typical signal\--to\--noise ratio of the final spectra is $\geq$30. Figure~\ref{specex} shows two stars with 
$T_{eff}$=3200K and 3400K, and Solar metallicity.

The same spectral analysis employed for our Baade's window and the (l, b)=(0$^\circ$, -1.0$^\circ$) fields 
\citep{rich05,rich07} has been performed here. 
A grid of suitable synthetic spectra of giant stars have been computed by varying
the photospheric parameters and the element abundances, using the code 
and line lists described in \citet{ori02}. 
The code uses the LTE approximation and is based on the molecular blanketed model atmospheres of \citet{jbk80} in the 3000-4000~K temperature range
and the ATLAS9 models for temperatures above 4000~K.
Very similar results have been obtained using the NextGen model atmospheres 
by \citet{hau99} \citep[see ][for more details]{ori02}. 
This is not surprising, since in the near IR the major source of continuum opacity is 
H$^-$ with its minimum near 1.6 $\mu$m, the dependence of the results on the choice 
of reasonable model atmospheres is not critical.

By combining full spectral synthesis analysis with
equivalent width measurements of selected lines, we derived abundances for Fe, C, O and
other $\alpha$\--elements (Mg, Si, Ca, and Ti). 
Reference solar abundances are from \citet{gs98}.
The lines and analysis method
have been subjected to rigorous tests in our previous 
studies of Galactic bulge field and cluster giants \citep[see][and references therein]{ori04,rich05,rich07,val11}.  We recall that the CO and especially OH bands  are extremely sensitive to $3500K<T_{eff}<4500K$.
The effective temperature determines the fraction of molecular {\it versus} atomic carbon and oxygen.
At temperatures $\ge$4500 K molecules barely survive; most of the carbon and oxygen 
are in atomic form and the CO and OH spectral features become very weak.
At temperatures $\le$3500 K most of the carbon and oxygen
are in molecular form, drastically reducing the dependence of the CO and OH
band strengths and equivalent widths on the temperature itself \citep{ori97}. 
At spectral resolution R$\sim$25,000, models with $\pm$0.2~dex abundance or 
$\pm$200~K temperature variations give remarkably  different molecular line profiles.
A microturbulence variation of $\pm$0.5~km/s mainly affects OH lines, while
gravity mainly affects CO lines.

In the first iteration, we estimate the stellar temperatures from the (J \--- K)$_0$ colors 
(see Tables~\ref{tabF175} and \ref{tabF265}) and the color temperature transformation of 
\citet{mont98} specifically
calibrated on globular cluster giants. Gravity has been estimated from theoretical evolutionary
tracks ($\log g \approx$0.5), according to the location of the stars on the RGB 
\citep[see][and references therein for a more detailed discussion]{ori97}. 
An average value $\xi$=2.0~kms$^{-1}$ has been adopted for the microturbulence \citep[see also][]{ori97}.
More stringent constraints on the stellar parameters are obtained by the simultaneous spectral fitting of
the several CO and OH molecular bands, which are very sensitive to the temperature, gravity and
microturbulence variations \citep[see Figs.~6 and 7 of][]{ori02}.

The final values of our best\--fit [Fe/H], [Si/Fe], [Mg/Fe], [O/Fe], [Ti/Fe], [C/Fe] and 
$\rm ^{12}$C/$^{13}$C abundances and abundance ratios together with their
random observational errors (on average $\pm$0.1 dex) are listed in Tables~\ref{tabF175} and \ref{tabF265}, for the 
F175 and F265 fields, respectively.
We also estimate a conservative $\leq$0.1~dex systematic error in the derived best\--fit abundances, 
due to residual uncertainty in the adopted
stellar parameters. However, it must be stressed that since the stellar features under 
consideration show a 
similar trend with variations in the stellar parameters, although with different sensitivities \citep{ori02}, {\it relative}
abundances are less dependent on the adopted stellar parameters (i.e. on the systematic errors) and their 
values are well constrained down to $\approx \pm$0.1~dex (see also Tables~\ref{tabF175} and \ref{tabF265}).
Tables~\ref{tabF175} and \ref{tabF265} also list the derived radial velocities for the stars in our sample.

\section{Results and Discussion}

From the abundances and abundance ratios reported in
Tables~\ref{tabF175} and \ref{tabF265} one determines the following average values 
and 1$\sigma$ errors for the  
F175 field: 
$\rm \langle [Fe/H] \rangle = -0.16 \pm 0.12$,
$\rm \langle [Ca/Fe] \rangle = +0.32 \pm 0.05$,
$\rm \langle [Si/Fe] \rangle = +0.30 \pm 0.07$,
$\rm \langle [Mg/Fe] \rangle = +0.32 \pm 0.06$,
$\rm \langle [Ti/Fe] \rangle = +0.30 \pm 0.09$,
$\rm \langle [O/Fe] \rangle = +0.34 \pm 0.05$.

For the F265 field, we find: 
$\rm \langle [Fe/H] \rangle = -0.21 \pm 0.08$,
$\rm \langle [Ca/Fe] \rangle = +0.30 \pm 0.04$,
$\rm \langle [Si/Fe] \rangle = +0.26 \pm 0.07$,
$\rm \langle [Mg/Fe] \rangle = +0.30 \pm 0.04$,
$\rm \langle [Ti/Fe] \rangle = +0.25 \pm 0.06$,
$\rm \langle [O/Fe] \rangle = +0.30 \pm 0.07$.

Figure~\ref{istofe} shows the iron abundance distributions for the four observed fields and the global sample
histogram superimposed, which gives $\rm <[Fe/H]>=-0.20\pm 0.02$ and 1$\sigma$ dispersion of 
0.11$\pm $0.01. 
We confirm the lack of substantial vertical iron abundance gradients from the center 
out to the Baade's window found by \citet{rich07}.
Our M giant [Fe/H] distributions peak at about the same metallicity as the [Fe/H] distributions 
of red clump stars published in the last few years and based on optical spectroscopic surveys of several 
tens to several hundred objects in Baade's window and in other outer bulge fields. 
At variance with these distributions, we do not find both 
{\it metal poor} ([Fe/H]$<-0.7$) and  {\it super-solar} ([Fe/H]$>0.2$) M giants.
The absence of metal poor stars is less of a concern, since
we select cool red giants, and globular-cluster age stars with [Fe/H]$\le-1$ 
 do not become M giants \citep{mm78,fer2000}.
However, the lack of significantly {\it super-solar} 
M giants is somewhat of a concern, although our finding is consistent with other small surveys of M giants and 
supergiants in the center of the Galaxy \citep{ram00b,cun06,cun07}, where no M stars with iron abundance of 
twice or three times solar have been measured.  
Further, we are consistent with an earlier medium resolution (R=4000) infrared study of M giants in the inner bulge, 
in field locations similar to ours \citep{ram00a}.
Indeed, although super-solar stars represent only a fraction of the stellar population towards the bulge,
they are definitely present in all of the optical studies based on red clump and K giants.   
We explored attempting to force a metal rich model (+0.3 dex) to fit the spectra; this failed at $>4\sigma$.   
We note that we have used this identical approach to measure more metal rich stars, reaching [Fe/H]=+0.4 in NGC 6791 and Terzan 5 \citep{n67,ter5}; the NGC 6791 result concurs with optical high resolution spectroscopy of its giants \citep{gra06}.

\begin{figure}
\plotone{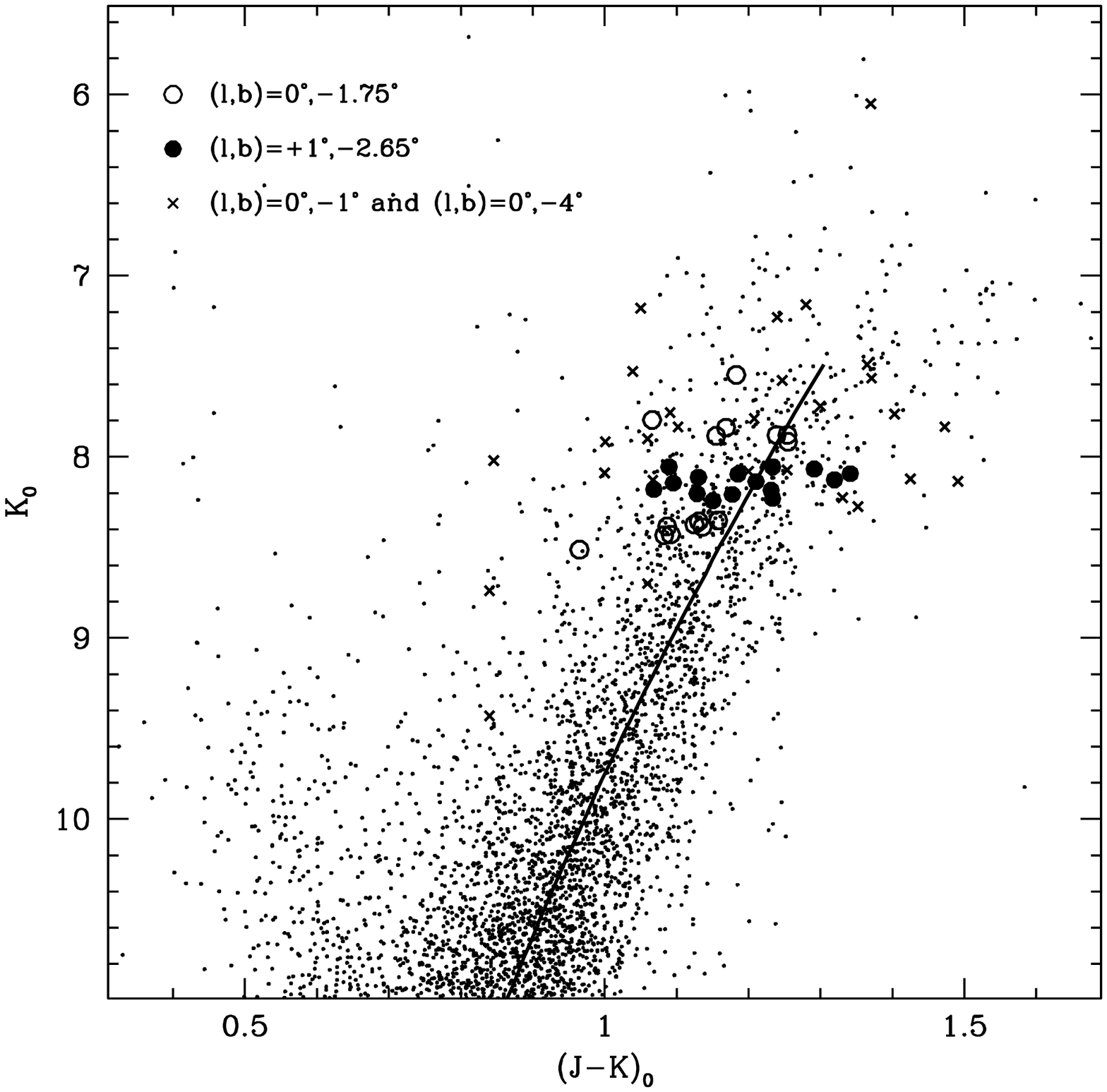}
\caption{2MASS K$_0$,(J-K)$_0$ color-magnitude diagram of the Baade's window, dereddened according to the prescription in 
\citet{gon11ab}. Superimposed are the M giants surveyed with NIRSPEC by our group in the Baade's window itself, and in other three inner bulge 
fields, namely F175 and F265 (this work) and the innermost field at b=-1$^{\circ}$ \citep{rich07}. We also superimpose the mean RGB ridge line of NGC~6528, a metal rich bulge cluster, for comparison.}
\label{cmd}   
\end{figure}

\begin{figure}
\plotone{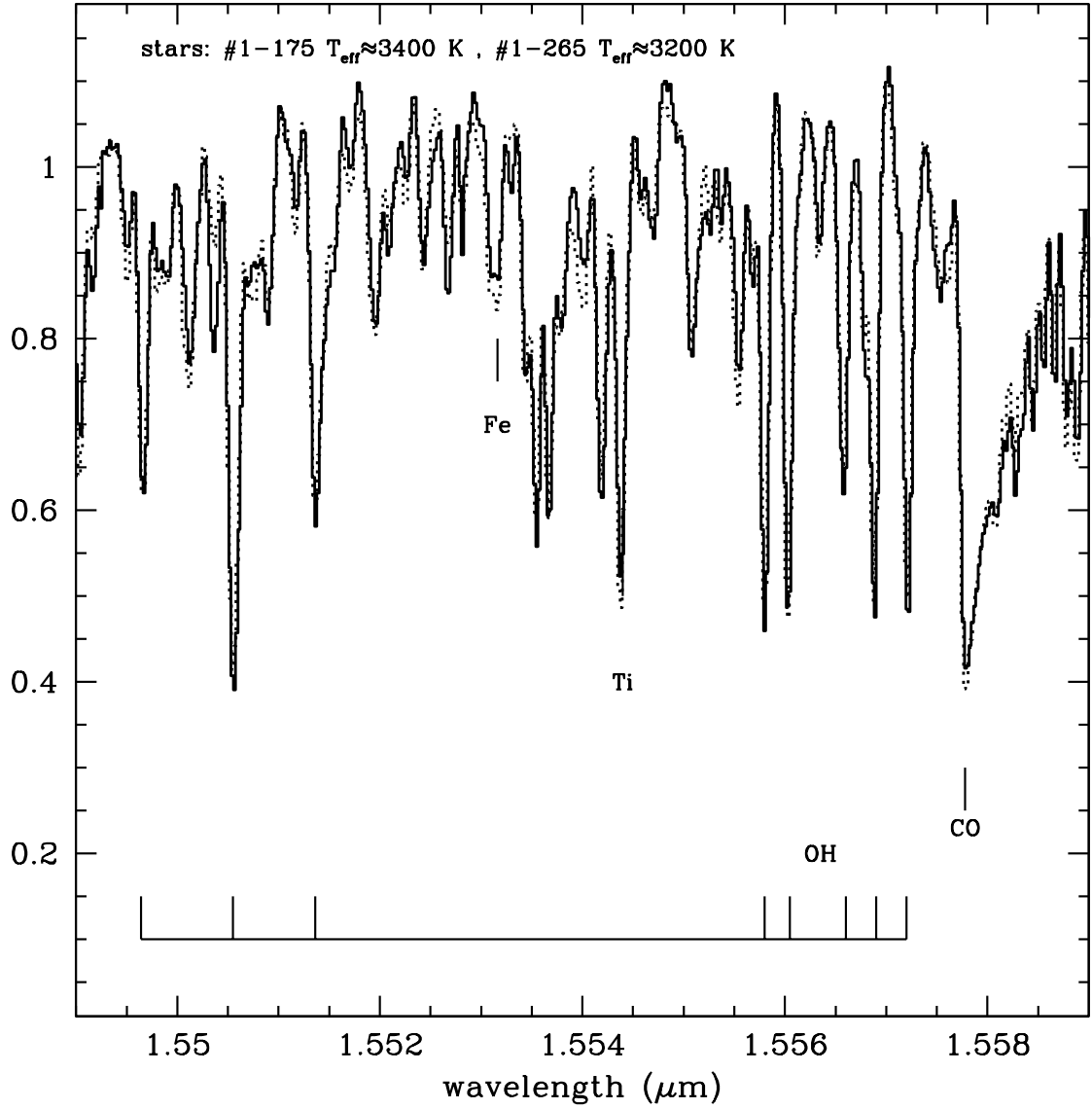}
\caption{NIRSPEC H\--band spectrum showing the region near 1.555 $\mu$ m of two (stars 1\--175 and 1\--265)
of the coolest and more metal\--rich  stars in our sample. A few major atomic lines and molecular bands of
interest are flagged.}
\label{specex}
\end{figure}

\begin{figure}
\plotone{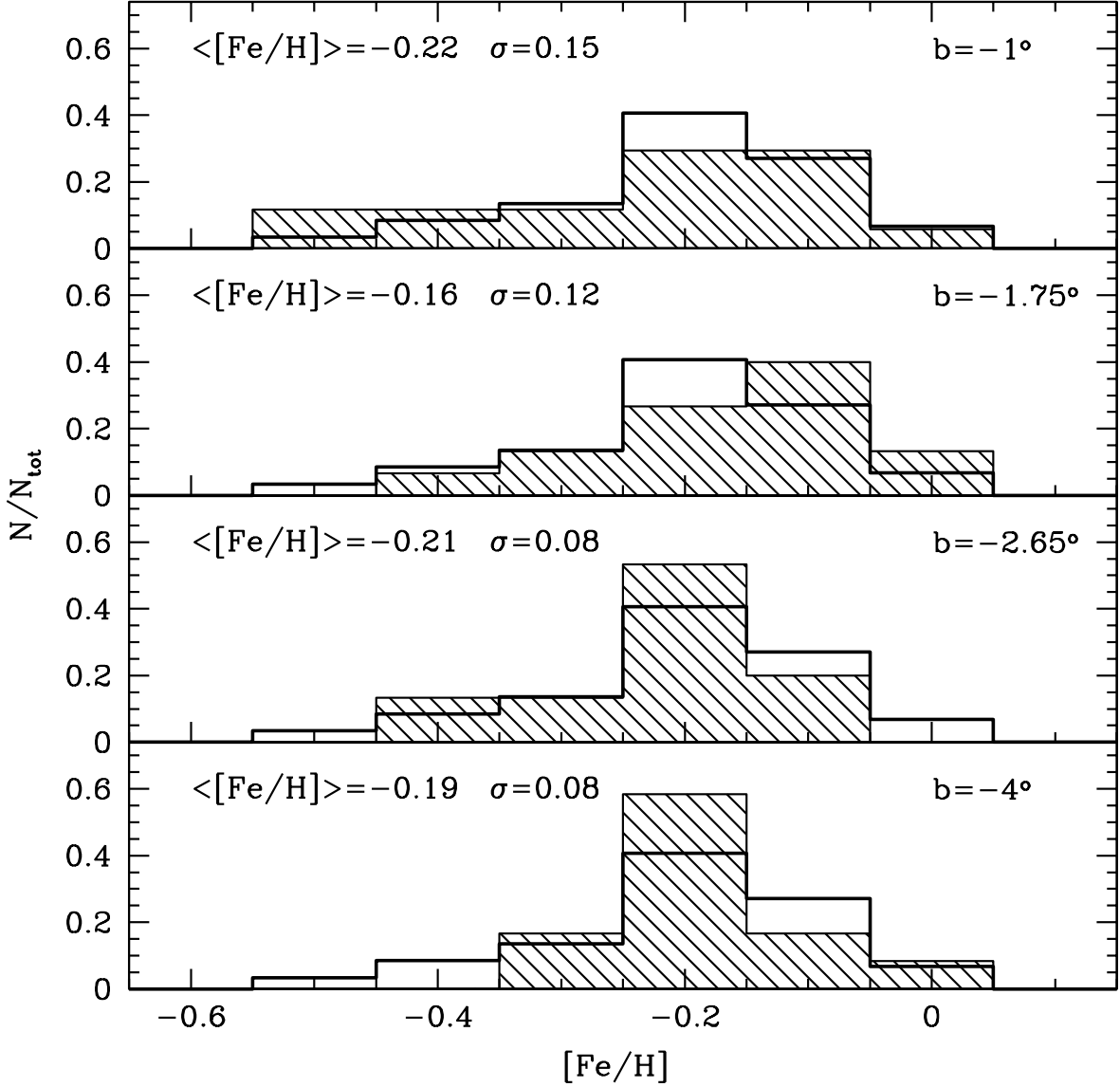}
\caption{Histograms (shaded) of the [Fe/H] distribution for the observed giants in the bulge fields at
(l,b)=(0$^\circ$, -1$^\circ$) ({\it top panel}, from \citet{rich07}), 
(l,b)=(0$^\circ$, -1.75$^\circ$) and 
(l,b)=(1$^\circ$, -2.65$^\circ$) ({\it middle panels}, this work) and
in the Baade's window {\it bottom panel}, from \citet{rich05}).
The global sample histogram  (solid, thick line) is overplotted in each panel.}
\label{istofe}
\end{figure}

In principle, we cannot exclude completely the possibility of bias in our target selection. 
To reduce the risk of including nonmembers, field stars, and outliers, we only consider giants with (J-K)$_0<$1.4, as we do not know the reddest color that may be reached by the bulge red giants.  
However, the color swath in Figure~\ref{cmd} is generous and if significant numbers of super-solar stars were present, 
we believe that they should have been in our sample, based on our color selection boundary.

Also, more generally, the intrinsic small number statistics 
due to the fast evolutionary rate towards the RGB tip, particularly at very high metallicity 
\citep[as inferred form canonical evolutionary tracks by e.g.][]{pie06}, make
the detection of super-solar M giants less probable.
Finally, the possibly enhanced mass loss
at super solar metallicities \citep[see e.g.][]{cas93}
can also prevent the most metal rich stars from reaching the RGB tip.

\begin{figure}
\plotone{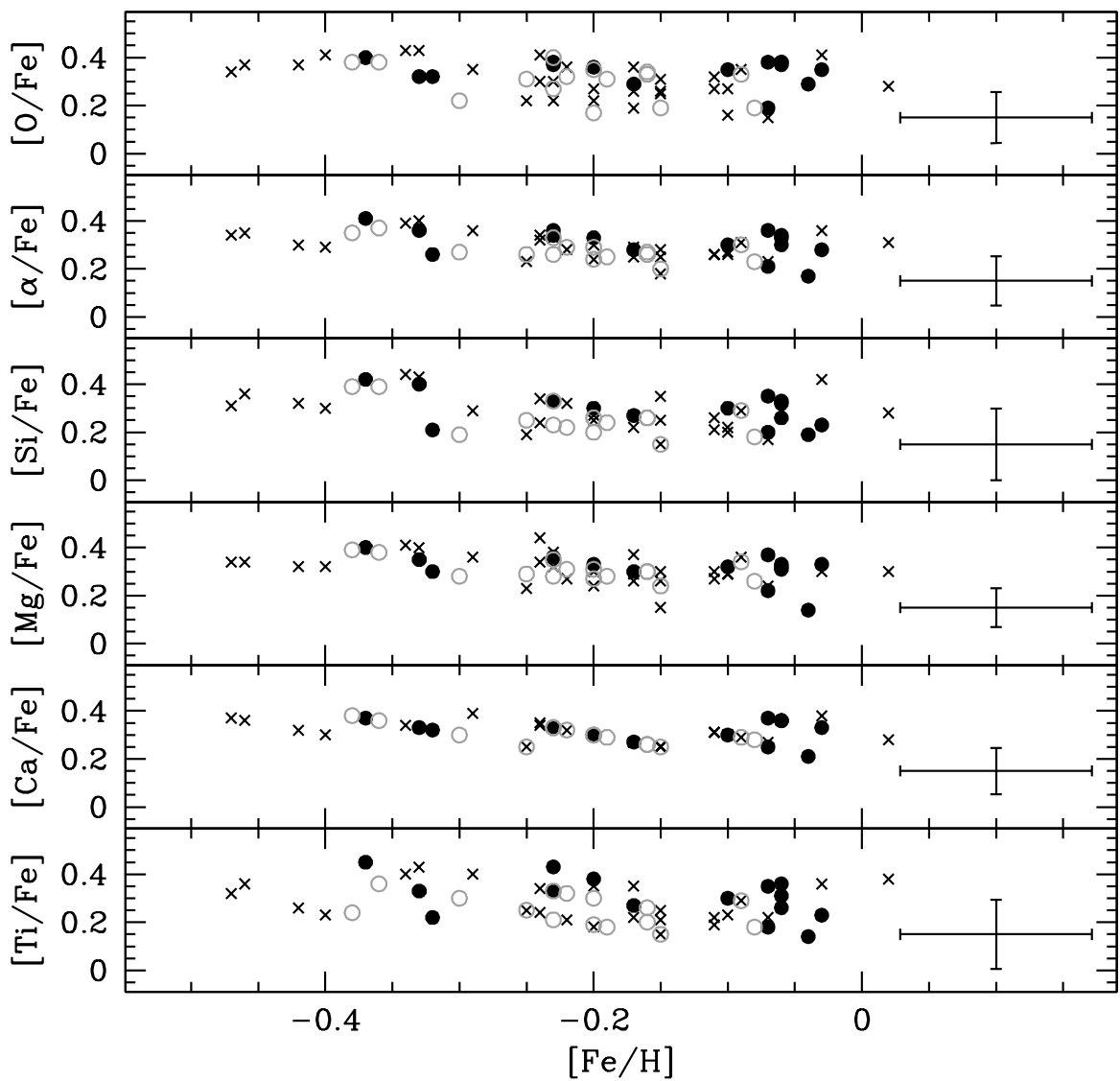}
\caption{[$\alpha$/Fe] abundance ratios as a function of [Fe/H] for the observed giants in the bulge fields at: 
(l, b)=(0$^\circ$, -1.75$^\circ$) (filled circles); (l,b)=(1$^\circ$, -2.65$^\circ$) (open circles);
(l,b)=(0$^\circ$, -1$^\circ$) and in the Baade's window (crosses, from \citep{rich05,rich07}, respectively).
Typical errors are plotted in the bottom corner of each panel.}
\label{alpha}
\end{figure}

\begin{figure}
\plotone{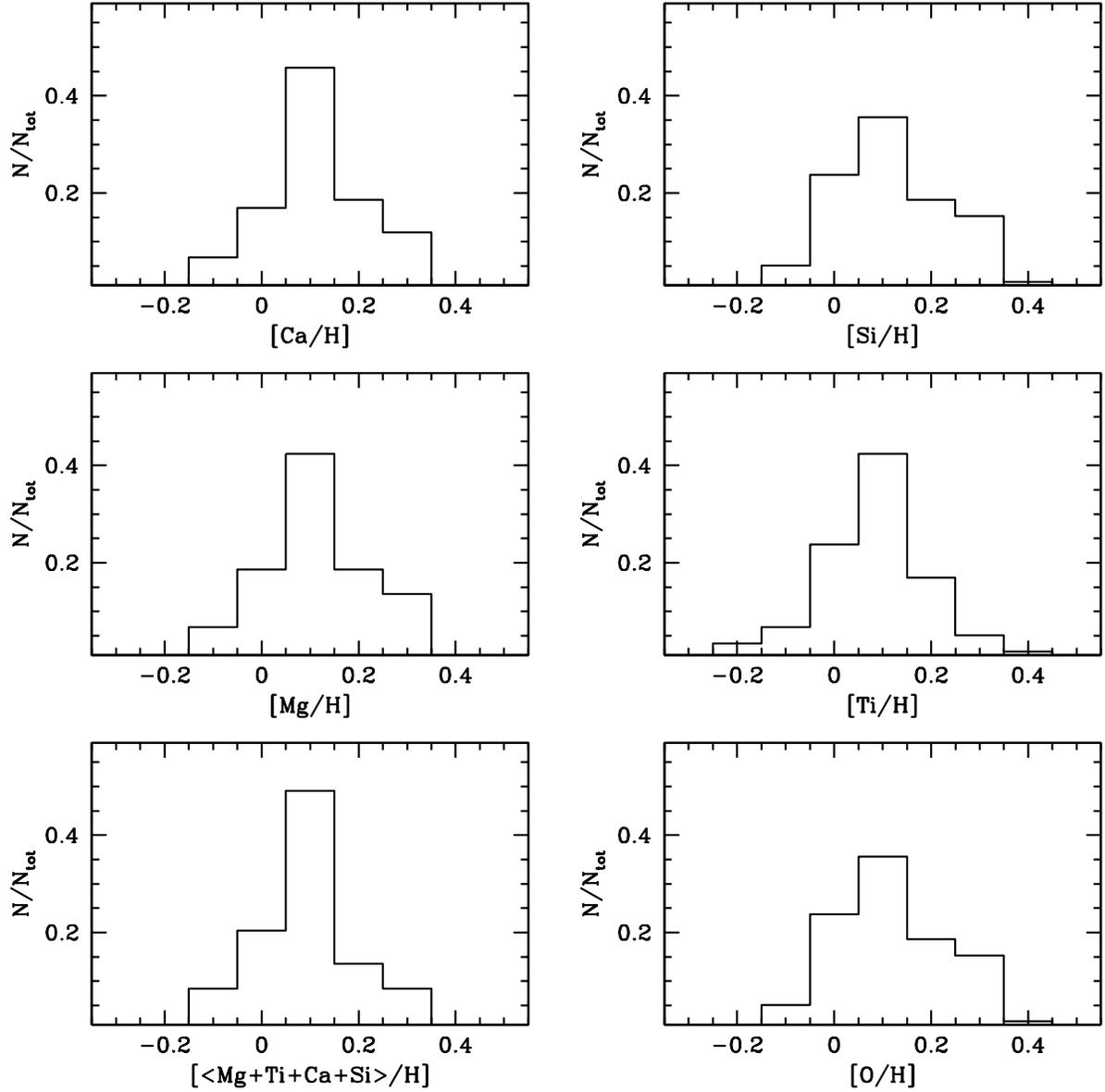}
\caption{Global sample [$\alpha$/H] abundance distributions for the 59 M giants in the four observed 
bulge fields.}
\label{istoalpha}
\end{figure}

Figure~\ref{alpha} shows the various [$\alpha$/Fe] abundance ratios as a function of the iron abundance 
for the M giants in the two inner fields presented in this paper and in the two previously studied fields,
for comparison, while Figure~\ref{istoalpha} shows the global sample histograms for the various 
[$\alpha$/H] abundances.

Figure~\ref{istoalpha} shows that M giants in all the four fields sampling the bulge stellar population in the innermost ~600 pc 
exhibit very homogeneous abundance patterns and an overall
constant [$\alpha$/Fe] enhancement at a level of $\approx$+0.3 dex up to about solar metallicity.  This suggests significant homogeneity to the enrichment process in the inner bulge.   
\citet{john11} and \citet{hill11} find a general pattern of alpha enhancement to nearly 1 kpc radius, but with a vertical gradient in [Fe/H].    Our study, which includes the sample of \citet{rich07}, and is consistent with \citet{ram00a}, finds no gradient in [Fe/H] or $\rm [\alpha/Fe]$ within $4^\circ$=550pc; this is in contrast with other spiral bulges that
have strong vertical Mg and Fe gradients in integrated optical light \citep{proc00}.  

Taking into consideration our results and those of \citet{john11}, \citet{hill11}, and \citet{gon11a}, we conclude that alpha enhancement is now shown to be characteristic of stars over the entire inner kpc of the bulge.   The apparent change in the [Fe/H] gradient from inner to outer bulge might on the surface appear to be consistent with the two-population picture of \citet{soto07} and \citet{hill11}: a metal rich but uniform bar at low latitude, transitioning into a metal poor classical spheroid at $r>500pc$.  However, the lack of kinematic evidence for the {\it metal poor} slowly rotating component in the {\it BRAVA} survey conflicts with the claim for a significant contingent, based on the bimodal deconvolution of the bulge abundance distribution by \citet{hill11}.

As in the previous studies of giant stars in the bulge field and globular clusters, 
we also find depleted [C/Fe] at the level of $\approx-0.3$ dex and 
low (5-10) $\rm ^{12}$C/$^{13}$C, indicative of extra-mixing processes during the 
evolution along the RGB. 

\begin{figure}
\plotone{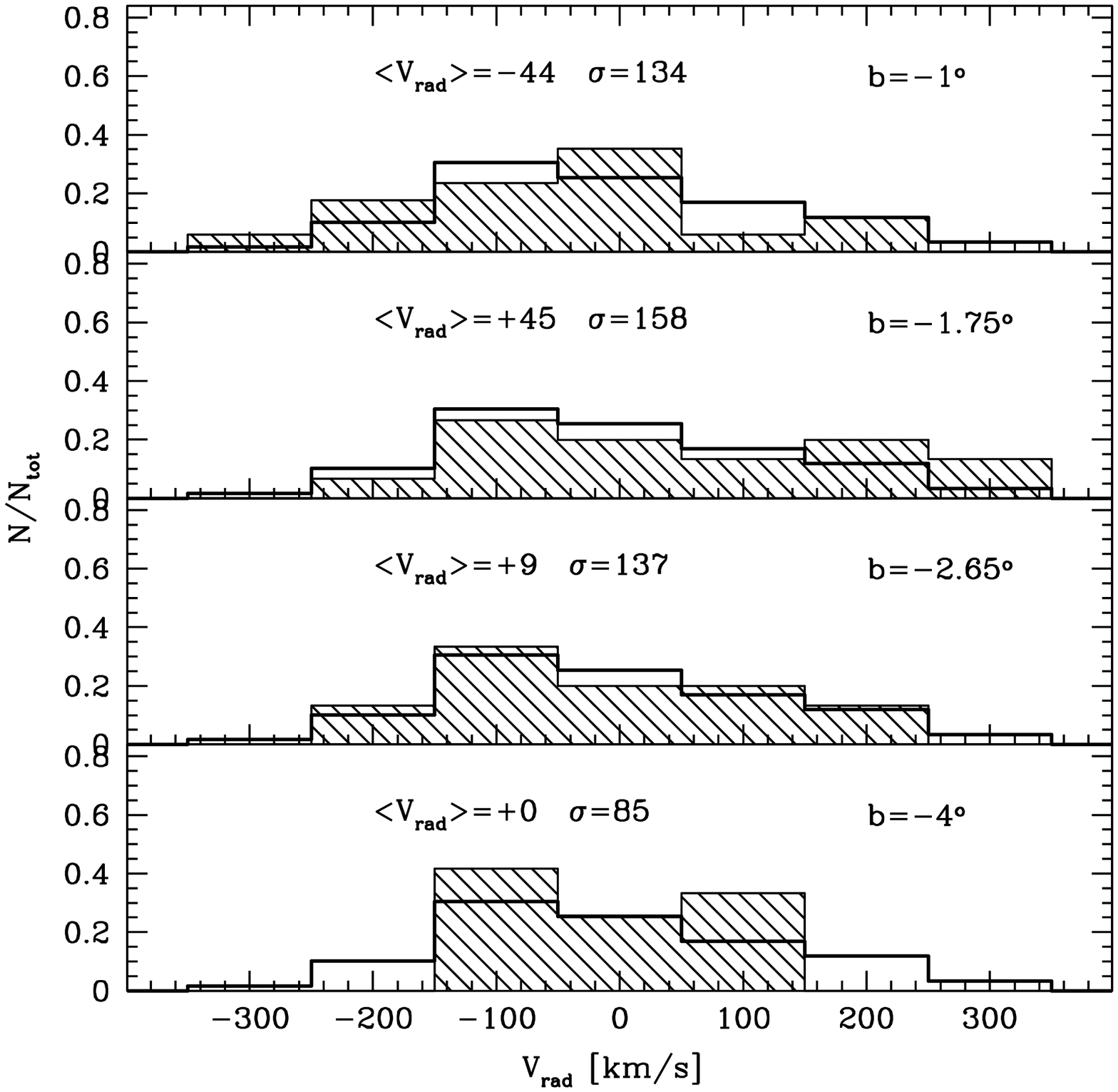}
\caption{Histograms (shaded) of the radial velocity distribution for the observed giants in the bulge fields at
(l,b)=(0$^\circ$, -1$^\circ$) ({\it top panel}, from \citet{rich07}), 
(l,b)=(0$^\circ$, -1.75$^\circ$) and 
(l,b)=(1$^\circ$, -2.65$^\circ$) ({\it middle panels}, this work) and
in the Baade's window ({\it bottom)}, from \citet{rich05}).
The global sample histogram (solid, thick line) is superimposed for each panel.  The global
dispersion of 134$~km\ sec^{-1}$ agrees with other bulge kinematics and also with the \citet{shen10} model; the 158$~km\ sec^{-1}$ dispersion in F175 is only $1\sigma$ high and would likely drop with increased sample size. }
\label{istovr}
\end{figure}

From the observed spectra we also measured radial velocities and velocity dispersion.
Figure~\ref{istovr} shows the radial velocity distributions for the four observed fields and the global sample histogram 
superimposed.
The overall velocity dispersion of the global sample of 59 M giants in the four fields is  $\rm \sigma_{v_r}\approx 134 \pm 12$ km/s,
while the mean radial velocity is +1.3$\pm$17 km/s, 
fully consistent with bulge kinematics from the {\it BRAVA} survey \citep{richetal07, how08, kun11} and consistent with the predictions
of the \citet{shen10} bar model.   We conclude that our sample of M giants in the inner bulge is
consistent with having kinematics expected for the $>$1 Gyr old bar/bulge population.

\acknowledgments

RMR acknowledges support from grant number 
AST-0709479 from the National Science Foundation. 

The authors are grateful to the staff at the Keck Observatory. 
The authors wish to recognize and acknowledge the very significant cultural role and reverence that the 
summit of Mauna Kea has always had within the indigenous Hawaiian community. We are most fortunate to 
have the opportunity to conduct observations from this mountain.

This paper is based upon work supported in part by the National 
Science Foundation under Grant No. 1066293 and the hospitality of the 
Aspen Center for Physics.

This publication makes use of data products from the Two Micron All Sky 
Survey (2MASS), which is a joint project of the University of 
Massachusetts and the Infrared Processing and Analysis Center/California 
Institute of Technology, founded by the National Aeronautics and Space 
Administration and the National Science Foundation.

The authors are grateful to an anonymous referee for insightful suggestions and comments.

\clearpage



\begin{deluxetable}{lccccccccccccc}
\tabletypesize{\scriptsize}
\rotate
\tablecaption{Stellar parameters and abundances for our giant stars in 
the F175 bulge field.\label{tabF175}}
\tablewidth{0pt}
\tablehead{
\colhead{Star} & 
\colhead{R.A.} &
\colhead{Decl} & 
\colhead{(J\--K)$_0^a$} &
\colhead{T$_{eff}$}&
\colhead{v$_r^b$} &
\colhead{[Fe/H]} &  
\colhead{[Ca/Fe]}&
\colhead{[Si/Fe]}& 
\colhead{[Mg/Fe]}&
\colhead{[O/Fe]}&
\colhead{[Ti/Fe]}&
\colhead{[C/Fe]}&
\colhead{$\rm ^{12}$C/$^{13}$C}\\
& (J2000.0) & (J2000.0)& & & (Km/s)& & & & &  & & &}
\startdata 
\multicolumn{14}{l}{}\\
1-175     &17:52:20.68   &-29:45:17.75  &1.255  &3400 &    29   &-0.03    &0.33  &0.23  &0.33	&0.35  &0.23	  & -0.27&  7.9\\
    &              &              &       &     &         &$\pm$0.04& $\pm$  0.06& $\pm$   0.19& $\pm$  0.06& $\pm$  0.10 & $\pm$    0.13& $\pm$      0.08 & $\pm$2.1\\
2-175     &17:52:15.52   &-29:44:26.23  &1.254  &3400 &  -111   &-0.06  &0.36  &0.32  &0.33   &0.37  &0.31	& -0.24&  6.3\\
    &              &              &       &     &         &$\pm$0.06& $\pm$   0.08& $\pm$   0.19& $\pm$  0.07& $\pm$  0.11 & $\pm$    0.14 & $\pm$     0.10 & $\pm$1.6\\
3-175     &17:52:15.07   &-29:44:31.42  &1.066  &3600 &     8   &-0.07  &0.37  &0.35  &0.37   &0.38  &0.35	& -0.23&  7.1\\
    &              &              &       &     &         &$\pm$0.08& $\pm$   0.09 & $\pm$  0.19& $\pm$  0.09& $\pm$  0.11 & $\pm$    0.17 & $\pm$     0.11 & $\pm$1.8\\
4-175     &17:52:21.15   &-29:45:59.01  &1.183  &3400 &   -93   &-0.06  &0.36  &0.26  &0.31   &0.38  &0.26	& -0.24&  8.9\\
    &              &              &       &     &         &$\pm$0.07& $\pm$   0.08& $\pm$   0.19& $\pm$  0.08& $\pm$  0.11 & $\pm$    0.14 & $\pm$     0.10& $\pm$2.3\\
5-175     &17:52:20.45   &-29:46:14.83  &1.155  &3600 &   274   &-0.06  &0.36  &0.33  &0.32   &0.38  &0.36	& -0.26&  6.3\\
    &              &              &       &     &         &$\pm$0.08& $\pm$   0.09& $\pm$   0.29& $\pm$  0.09& $\pm$  0.11 & $\pm$    0.17 & $\pm$     0.11 & $\pm$1.6\\
6-175     &17:52:06.70   &-29:58:41.79  &1.169  &3400 &   166   &-0.04  &0.21  &0.19  &0.14   &0.29  &0.14	& -0.36& 10.0\\
   &              &              &       &     &         &$\pm$0.05& $\pm$   0.07& $\pm$   0.18& $\pm$  0.06 & $\pm$ 0.10 & $\pm$    0.14 & $\pm$     0.08& $\pm$2.6\\
7-175     &17:52:13.16   &-29:58:15.85  &1.239  &3400 &    -9   &-0.07  &0.25  &0.20  &0.22   &0.19  &0.18	& -0.33&  7.1\\
    &              &              &       &     &         &$\pm$0.05& $\pm$   0.07& $\pm$   0.18& $\pm$  0.06& $\pm$  0.09 & $\pm$    0.14& $\pm$      0.08 & $\pm$1.8\\
8-175 	  &17:52:44.758  &-29:50:30.93  &1.158  &3400 &   119   &-0.17  &0.27  &0.27  &0.30   &0.29  &0.27	& -0.28&  6.3\\
    &              &              &       &     &         &$\pm$0.08& $\pm$   0.12& $\pm$   0.27 & $\pm$ 0.09& $\pm$  0.11 & $\pm$    0.15 & $\pm$     0.11& $\pm$1.6\\
9-175 	  &17:52:18.860  &-29:43:38.98  &1.136  &3600 &   169   &-0.10  &0.30  &0.30  &0.32   &0.35  &0.30	& -0.30&  7.9\\
    &              &              &       &     &         &$\pm$0.08& $\pm$   0.09 & $\pm$  0.28& $\pm$  0.09& $\pm$  0.11 & $\pm$    0.14 & $\pm$     0.11& $\pm$ 2.1\\
10-175 	  &17:52:39.456  &-29:50:39.96  &1.131  &3600 &  -111   &-0.20  &0.30  &0.30  &0.33   &0.36  &0.38	& -0.25&  4.5\\
    &              &              &       &     &         &$\pm$0.08& $\pm$   0.12& $\pm$   0.27& $\pm$  0.08& $\pm$  0.10& $\pm$     0.14 & $\pm$     0.10 & $\pm$1.2\\
11-175	  &17:52:43.323  &-29:59:43.95  &1.125  &3600 &    73   &-0.33  &0.33  &0.40  &0.35   &0.32  &0.33	& -0.47&  7.9\\
    &              &              &       &     &         &$\pm$0.08& $\pm$   0.11 & $\pm$  0.21& $\pm$  0.09& $\pm$  0.11 & $\pm$    0.14 & $\pm$     0.11& $\pm$ 2.1\\
12-175	  &17:52:40.730  &-29:59:42.63  &1.092  &3600 &   -87   &-0.23  &0.33  &0.33  &0.36   &0.38  &0.33	& -0.27&  6.3\\
    &              &              &       &     &         &$\pm$0.08& $\pm$   0.12& $\pm$   0.27& $\pm$  0.08& $\pm$  0.11 & $\pm$    0.14& $\pm$      0.10& $\pm$1.6\\
13-175	  &17:52:58.997  &-29:38:50.00  &1.087  &3600 &   284   &-0.23  &0.33  &0.33  &0.36   &0.37  &0.43	& -0.22&  6.3\\
    &              &              &       &     &         &$\pm$0.07& $\pm$   0.12& $\pm$   0.27& $\pm$  0.08& $\pm$  0.11& $\pm$     0.14& $\pm$      0.10 & $\pm$1.6\\
14-175	  &17:52:42.627  &-29:54:11.25  &1.083  &3600 &  -242   &-0.32  &0.32  &0.21  &0.30   &0.32  &0.22	& -0.38&  5.0\\
    &              &              &       &     &         &$\pm$0.07& $\pm$   0.10& $\pm$   0.23& $\pm$  0.07& $\pm$  0.10& $\pm$     0.13& $\pm$      0.10 & $\pm$1.3\\
15-175	  &17:52:53.328  &-29:58:49.84  &0.965  &3800 &   220   &-0.37  &0.37  &0.42  &0.40   &0.40  &0.45	& -0.23&  6.3\\ 
   &              &              &       &     &         &$\pm$0.10& $\pm$   0.13& $\pm$   0.23& $\pm$  0.11& $\pm$  0.11& $\pm$     0.15& $\pm$      0.12& $\pm$ 1.6\\
\enddata
\tablenotetext{a}{The (J\--K) colors are from 2MASS and have been corrected for reddening using E(B \--V)=1.2}
\tablenotetext{b}{Heliocentric radial velocity in km/s.}
\end{deluxetable}
\clearpage

\begin{deluxetable}{lccccccccccccc}
\tabletypesize{\scriptsize}
\rotate
\tablecaption{Stellar parameters and abundances for our giant stars in 
the F265 bulge field.\label{tabF265}}
\tablewidth{0pt}
\tablehead{
\colhead{Star} & 
\colhead{R.A.} &
\colhead{Decl} & 
\colhead{(J\--K)$_0^a$} &
\colhead{T$_{eff}$}&
\colhead{v$_r^b$} &
\colhead{[Fe/H]} &  
\colhead{[Ca/Fe]}&
\colhead{[Si/Fe]}& 
\colhead{[Mg/Fe]}&
\colhead{[O/Fe]}&
\colhead{[Ti/Fe]}&
\colhead{[C/Fe]}&
\colhead{$^{12}$C/$^{13}$C}\\
& (J2000.0) & (J2000.0)& & & (Km/s)& & & & &  & & &}
\startdata 
\multicolumn{14}{l}{}\\
1-265	 &17:58:37.119  &-29:03:48.39   &1.346  &3200&   -96 &-0.08  &0.28  &0.18   &0.26  &0.19   &0.18 &-0.32 &  7.9\\
    &              &              &       &     &         &$\pm$0.07&$\pm$   0.08&$\pm$   0.17&$\pm$   0.08&$\pm$  0.12&$\pm$   0.13&$\pm$  0.10&$\pm$  2.1\\
2-265	 &17:58:49.636  &-29:23:32.65   &1.324  &3200&   -12 &-0.15  &0.25  &0.15   &0.24  &0.19   &0.15 &-0.38 & 10.0\\
    &              &              &       &     &         &$\pm$0.06&$\pm$   0.11&$\pm$   0.16&$\pm$   0.07&$\pm$  0.10&$\pm$   0.14&$\pm$  0.09&$\pm$  2.6\\ 
3-265	 &17:58:43.132  &-29:18:35.92   &1.296  &3200&   238 &-0.20  &0.30  &0.20   &0.27  &0.17   &0.19 &-0.30 &  8.9\\
    &              &              &       &     &         &$\pm$0.10 &$\pm$  0.13&$\pm$   0.17&$\pm$   0.10&$\pm$  0.14&$\pm$   0.16&$\pm$  0.12&$\pm$  2.3\\  
4-265	 &17:58:31.816  &-29:19:24.17   &1.238  &3400&  -112 &-0.22  &0.32  &0.22   &0.31  &0.32   &0.32 &-0.26 & 10.0\\
    &              &              &       &     &         &$\pm$0.06&$\pm$   0.11&$\pm$   0.17&$\pm$   0.07&$\pm$  0.10&$\pm$   0.15&$\pm$  0.09&$\pm$  2.6\\
5-265	 &17:58:35.654  &-29:04:52.58   &1.238  &3400&   147 &-0.09  &0.29  &0.29   &0.34  &0.33   &0.29 &-0.31 &  7.9\\
    &              &              &       &     &         &$\pm$0.05&$\pm$   0.07&$\pm$   0.18 &$\pm$  0.06&$\pm$  0.10&$\pm$   0.14&$\pm$  0.08&$\pm$  2.1\\
6-265	 &17:58:51.445  &-29:18:08.60   &1.236  &3400&   136 &-0.30  &0.30  &0.19   &0.28  &0.22   &0.30 &-0.25 & 10.0\\
    &              &              &       &     &         &$\pm$0.08&$\pm$   0.12&$\pm$   0.17 &$\pm$  0.08&$\pm$  0.11&$\pm$   0.14&$\pm$  0.10&$\pm$  2.6\\
7-265	 &17:58:35.039  &-29:02:05.51   &1.215  &3400&   -81 &-0.23  &0.33  &0.23   &0.28  &0.27   &0.21 &-0.27 &  8.9\\
    &              &              &       &     &         &$\pm$0.08&$\pm$   0.12&$\pm$   0.18 &$\pm$  0.09&$\pm$  0.11&$\pm$   0.15&$\pm$  0.11&$\pm$  2.3\\
8-265	 &17:58:43.835  &-29:07:42.54   &1.190  &3400&   -97 &-0.19  &0.29  &0.24   &0.28  &0.31   &0.18 &-0.28 &  8.9\\
    &              &              &       &     &         &$\pm$0.06&$\pm$   0.11&$\pm$   0.17&$\pm$   0.07&$\pm$  0.10&$\pm$   0.15&$\pm$  0.10&$\pm$  2.3\\  
9-265	 &17:58:55.474  &-29:22:12.42   &1.182  &3400&  -151 &-0.25  &0.25  &0.25   &0.29  &0.31   &0.25 &-0.37 & 10.0\\
    &              &              &       &     &         &$\pm$0.08&$\pm$   0.12&$\pm$   0.17&$\pm$   0.08&$\pm$  0.11&$\pm$   0.14&$\pm$  0.10&$\pm$  2.6\\
10-265	 &17:58:36.306  &-29:10:38.64   &1.155  &3600&     6 &-0.16  &0.26  &0.26   &0.30  &0.34   &0.26 &-0.34 &  6.3\\
    &              &              &       &     &         &$\pm$0.07&$\pm$   0.11&$\pm$   0.17 &$\pm$  0.08&$\pm$  0.10&$\pm$   0.13&$\pm$  0.10&$\pm$  1.6\\
11-265	 &17:58:33.611  &-29:02:22.20   &1.135  &3600&     7 &-0.23  &0.33  &0.33   &0.35  &0.40   &0.33 &-0.24 &  7.9\\
    &              &              &       &     &         &$\pm$0.06&$\pm$   0.11&$\pm$   0.17&$\pm$   0.07&$\pm$  0.1&$\pm$0   0.13&$\pm$  0.09&$\pm$  2.1\\
12-265	 &17:58:49.028  &-29:11:22.60   &1.133  &3600&   121 &-0.16  &0.26  &0.26   &0.30  &0.33   &0.20 &-0.44 &  8.9\\
    &              &              &       &     &         &$\pm$0.07&$\pm$   0.11&$\pm$   0.17&$\pm$   0.08&$\pm$  0.09 &$\pm$  0.13&$\pm$  0.10&$\pm$  2.3\\
13-265	 &17:58:37.405  &-28:59:33.23   &1.100  &3600&   -65 &-0.36  &0.36  &0.39   &0.38  &0.38   &0.36 &-0.29 & 10.0\\
    &              &              &       &     &         &$\pm$0.07&$\pm$   0.10&$\pm$   0.18 &$\pm$  0.07&$\pm$  0.10&$\pm$   0.13 &$\pm$ 0.10&$\pm$  2.6\\   
14-265	 &17:58:55.408  &-29:07:07.30   &1.094  &3600&   250 &-0.20  &0.30  &0.26   &0.31  &0.35   &0.30 &-0.30 &  7.1\\
    &              &              &       &     &         &$\pm$0.08&$\pm$   0.12&$\pm$   0.17&$\pm$   0.08&$\pm$  0.11&$\pm$   0.14&$\pm$  0.10&$\pm$  1.8\\ 
15-265	 &17:58:43.660  &-29:03:25.69   &1.073  &3600&  -156 &-0.38  &0.38  &0.39   &0.39  &0.38   &0.24 &-0.42 &  8.9\\
    &              &              &       &     &         &$\pm$0.07&$\pm$   0.10&$\pm$   0.19&$\pm$   0.07&$\pm$  0.10&$\pm$   0.13&$\pm$  0.10&$\pm$  2.3\\    
\enddata
\tablenotetext{a}{The (J\--K) colors are from a catalog obtained by matching 2MASS J band with LCO K band. A reddening E(B\--V)=0.65 has been used.}
\tablenotetext{b}{Heliocentric radial velocity in km/s.}
\end{deluxetable}
\clearpage


\clearpage

\end{document}